\documentclass[aip,reprint, graphicx]{revtex4-1}
\usepackage{graphicx}
\usepackage{amsmath}
\usepackage{xr}
\usepackage[version=4]{mhchem}
\usepackage{subcaption}
\usepackage{bm}% bold math
\usepackage[utf8]{inputenc}
\usepackage[T1]{fontenc}
\usepackage{mathptmx}
\usepackage{etoolbox}

\usepackage{dcolumn}
\draft
\makeatletter
\def\@email#1#2{%
 \endgroup
 \patchcmd{\titleblock@produce}
  {\frontmatter@RRAPformat}
  {\frontmatter@RRAPformat{\produce@RRAP{*#1\href{mailto:#2}{#2}}}\frontmatter@RRAPformat}
  {}{}
}%
\makeatother
\begin{document}

\title{A Direct Diabatic States Construction Method with Consistent Orbitals for Valence and Rydberg States}
% Force line breaks with \\

\author{Jiamin Jin}
\affiliation{Institute of Theoretical and Computational Chemistry, Key Laboratory of Mesoscopic Chemistry of the Ministry of Education (MOE), School of Chemistry and Chemical Engineering, Nanjing University, Nanjing 210023, China}

\author{Zexing Qu}
\email{zxqu@jlu.edu.cn}
\affiliation{Institute of Theoretical Chemistry, College of Chemistry, Jilin University, Changchun 130023, China}

\author{Chungen Liu}
\email{cgliu@nju.edu.cn}
\affiliation{Institute of Theoretical and Computational Chemistry, Key Laboratory of Mesoscopic Chemistry of the Ministry of Education (MOE), School of Chemistry and Chemical Engineering, Nanjing University, Nanjing 210023, China}

\begin{abstract}
This work presents a novel methodology termed Direct Diabatic States Construction (DDSC), which integrates fragment wavefunctions into an anti-symmetric wavefunction for the entire system. Using a fragment-localized state-consistent molecular orbital (FL-SC MO), this approach enables direct construction of all diabatic states at the same root. Each diabatic state is formed as a linear combination of a set of diabatic configurations. The validity and effectiveness of DDSC have been demonstrated through its application to the LiH molecule. The results show that this method is suitable for constructing both valence and Rydberg diabatic states. One of the key advantages of DDSC is its ability to directly compute diabatic couplings, which can be converted to non-adiabatic coupling (NAC) vectors along the reaction coordinate. The DDSC method efficiently builds the diabatic potential energy matrix (DPEM), especially for systems with clear fragment partitions and weak inter-fragment interactions, such as charge transfer reactions.
\end{abstract}

\maketitle

\section{\label{sec:level1}Introduction}
The Born-Oppenheimer (BO) approximation constitutes a cornerstone in modern quantum chemistry, providing the foundation for generating adiabatic potential energy surfaces (PESs) that are essential for exploring the mechanisms of chemical reactions. Adiabatic processes can be effectively simulated using BO molecular dynamics (BOMD) on a single PES. However, certain processes, such as photo-dissociation, charge transfer, and energy transfer, face challenges due to the proximity of multiple adiabatic PESs, specifically in regions identified as avoided crossings or conical intersections (CIs) \cite{Yarkony1996, Yarkony1996a, Yarkony1998}. In these areas, nuclear and electronic motions are strongly coupled, leading to a sharp increase in non-adiabatic couplings (NACs). Accurate computation of NACs requires advanced quantum chemical methods because of the multiconfigurational nature of the electronic wavefunctions. To address these complexities, the diabatic representation offers a viable alternative. In this framework, the NACs are eliminated, albeit introducing electronic couplings between different diabatic states \cite{Jasper2006, Kouppel1984, Baer2002, Worth2004, Shu2022, Tannor2007}. Diabatic wavefunctions, when properly constructed, exhibit slow variation with respect to molecular geometry, even in the vicinity of CIs. The diabatic potential energy surfaces (PESs) and diabatic couplings are smooth functions of the reaction coordinates, which facilitates their analytical representation. This property is particularly beneficial for the development of theoretical methodologies designed to investigate non-adiabatic effects in large molecular systems \cite{Shu2022, VanVoorhis2010, Wang2022, Qiu2022}.

It has been theoretically established that strictly defined diabatic states do not exist in polyatomic systems\cite{Mead1982}. Despite this limitation, \textit{quasi}-diabatic states can be effectively constructed through either adiabatic-to-diabatic (ATD) transformation techniques or direct construction methodologies. The ATD approaches primarily pursue transformations through mathematical minimization of non-adiabatic couplings (NACs) \cite{Abrol2002, Baer1975}, maximization of differences in specific physical properties between distinct electronic states \cite{Levy1992, Pettersson2006, Hsu2009, Hoyer2014, Hoyer2016}, or enhancement of configuration consistency in each diabatic state. \cite{Pacher1988, Atchity1997, Ruedenberg1993, Nakamura2001, Nakamura2002}. 
Despite their massive outstanding achievements, the ATD framework can be computationally demanding due to the necessity of calculating multiconfigurational electronic wavefunctions, as is required in the adiabatic representation framework. Alternatively, direct construction methods can represent each diabatic state with specially designed electronic configurations, which can be computed with less expensive quantum chemical methods. Several strategies are available to build such states, including the constrained density functional theory (CDFT) \cite{Wu2005, Wu2006, Wu2006a, Wu2007,Kaduk2012, Holmberg2017, Holmberg2018}, Valence Bond (VB) theories \cite{Wu2011, Su2013, Lin2018, Zhang2020b}, and multistate density functional theory (MSDFT)\cite{Cembran2009, Gao2016, Grofe2017, Liu2018a}. In CDFT, diabatic states are optimized through imposing electron density constraints on predefined fragments in the system. This approach is suitable for electron transfer reactions, while it faces challenges in characterizing the local excitation configurations. Moreover, the schemes of partitioning electron density distribution \cite{Meng2022}, as well as the choice of exchange-correlation functionals \cite{Mavros2015} can significantly impact the accuracy of the diabatic states. 
Compared to CDFT, VB theory enables the definition of a wider variety of configurations. A VB structure is expanded using a small number of selected configurations. This approach provides a natural description for the physical picture of the diabatic state \cite{VanVoorhis2010, Shu2022}, as a single VB structure can directly correspond to a diabatic state. Block-localized wavefunction (BLW)\cite{Mo2000, Song2008}, a modern extension of VB theory, employs fragment-localized molecular orbitals instead of atomic orbitals to simplify the definition of configurations. 
MSDFT \cite{Cembran2009, Gao2016, Grofe2017, Liu2018a} also adopts VB-type configurations to represent diabatic states. In MSDFT, electronic wavefunctions are constructed using Kohn-Sham fragment-localized orbitals. By incorporating dynamical correlations during orbital optimization, MSDFT offers a more efficient and accurate description of diabatic states than BLW-VB.

The molecular orbitals (MOs) employed in both BLW-VB and MSDFT methods are optimized individually for each diabatic state \cite{Song2008, Grofe2017}. Although these state-specific orbitals allow for a compact representation of distinct diabatic states, the inherent nonorthogonality between diabatic states resulting from these orbitals introduces considerable computational complexity when evaluating diabatic coupling matrix elements.
To address this issue, we propose a Direct Diabatic States Construction (DDSC) method, utilizing fragment-localized state-consistent molecular orbital (FL-SC MO), thereby eliminating inter-state molecular orbital nonorthogonality by design. The key advantages of this approach is its ability to compute diabatic couplings directly, which can be converted to non-adiabatic coupling vectors along the reaction coordinate. Furthermore, the present methodology enables convenient treatment of Rydberg states by precisely characterizing their unique electron occupation patterns within Rydberg orbitals, thereby establishing a unified theoretical framework for simultaneous description of both valence and Rydberg states. The characterization of the dissociation potential energy surfaces of LiH molecule will be used to demonstrate the methodology. LiH has been extensively investigated by numerous theoretical and experimental researchers \cite{Boutalib1992, Stwalley1993,Chen1999,Huang2000, Mo1993, Hoyer2016, Grofe2017, Liu2018a}. In LiH, it involves not only complicated interactions between ionic and covalent configurations in the valence states, but also the participation of Rydberg states in photo-excited dissociation processes. The weak Coulombic attraction of Li atom induces electron delocalization within its valence shell, which facilitates Rydberg state formation via electron diffusion, thereby introducing inherent complexity in theoretically characterizing the electronic structure.

\section{\label{sec:level1}Methodology}
In the DDSC approach, the molecular system was partitioned into fragments. Covalent and ionic configurations were created by the redistribution of electrons among these fragments. In the demonstrative study of LiH system, charged fragments (e.g. [Li\(^+\)], [Li\(^-\)], [H\(^+\)], [H\(^-\)]) and neutral ones (e.g. [Li\(\cdot\)] and [H\(\cdot\)]) were generated. Furthermore, local intra-fragment excitations could be described through the introduction of excitation configurations unique to each fragment. For example, the valence or Rydberg states of LiH were easily identified by relocating the 2\(s\) valence electron in Li atom ([Li(2\(s\))\(\cdot\)]) to the 2\(p\) orbital ([Li(2\(p\))\(\cdot\)]) or other more diffusive Rydberg orbitals ([Li(3\(s\))\(\cdot\)], [Li(3\(p\))\(\cdot\)]). The wavefunction of the entire system was derived from combining these fragment wavefunctions. The optimization of fragment orbitals as well as the construction of diabatic states are the two key points of our approach that will be detailed in the following paragraphs. Also, the diabatic-to-adiabatic (DTA) transformation and the calculation of non-adiabatic couplings will be introduced.

\subsection{\label{sec:level2}The optimization of fragment orbitals}
To generate FL-SC MO within the DDSC formalism, all diabatic states strictly maintained orbital invariance by employing a common orbital basis derived from a state-averaged reduced density matrix. This approach ensured an unbiased representation of both neutral and ionic configurations, providing a robust foundation for reproducing inter-state properties, such as energy gaps and avoided crossings in potential energy surfaces.

Our methodology employed a partitioned optimization strategy for fragment orbitals across individual subsystems, circumventing the computational overhead associated with multi-state electronic structure calculations for the complete molecular system. The protocol initiated with computations of fragment electronic states within the mean-field framework, incorporating configurations essential for resolving non-adiabatic processes. For each electronic state \(i\) localized on fragment \( M \), we formulated the first-order reduced density matrix (1-RDM)  \(\rho^i_M \). A configurationally averaged density matrix \(\mathbf{P}_M\) was subsequently constructed through the weighted superposition:  
\begin{equation}  
    \mathbf{P}_M = \sum_i w_i \rho^i_M,  
    \label{eq:rdm}  
\end{equation}  
where \(w_i\) denotes the configuration weighting factor. This procedure yielded fragment-specific orbitals through eigen-decomposition analysis of the transformed matrix product  \(\mathbf{P}_M\mathbf{S}_{\text{basis}} \), where  \(\mathbf{S}_{\text{basis}} \) represents the atomic orbital overlap matrix. The resulting eigenvectors were utilized as localized optimal molecular orbitals.

\subsection{\label{sec:level2}The construction of diabatic states }
The construction of individual diabatic states required a multiconfiguration expansion, based on the optimized fragment orbitals, to account for the state-specific wavefunction perturbations from the aspect of configuration interactions. Supposing the  \(i \) th configuration basis was represented with a Slater determinant  \(\Omega_i \)  created as the product of the Slater determinants from each fragment,
\begin{equation}
    \Omega_i =\hat{A}\prod_M D^{(i)}(M), 
\end{equation}
where  \(D^{(i)}(M) \) refers to the selected Slater determinant of fragment \(M\) for constructing configuration \(i\). An anti-symmetrization operator \(\hat{A}\) was introduced to ensure the antisymmetry of the total wavefunction. Notably, the combination of \(D^{(i)}(M)\) must adhere to the principle of electron number conservation of the composite system. To ensure proper spin symmetry and reduce the computational cost, one should linearly combine a series of spatially equivalent configuration basis functions to form configuration state functions (CSFs) through  
\begin{equation}  
    \Psi_i = \sum_k c_k^{(i)} \Omega_i^{(k)}.  
\end{equation}  
The expansion coefficients \(\{c_k^{(i)}\}\) were determined by requiring the wavefunction to be an eigenfunction of \(\mathbf{S}^2\).  

For diatomic molecule LiH, the diabatic CSFs were categorized into ionic (such as [Li\(^+\)][H\(^-\)] or [Li\(^-\)][H\(^+\)]), and covalent (such as [Li\(\cdot\)][H\(\cdot\)]) configurations. Additionally, they were further classified as valence and Rydberg configurations through specifying the occupied atomic orbitals (such as [Li(2\(s\))\(\cdot\)][H\(\cdot\)], [Li(2\(p\))\(\cdot\)][H\(\cdot\)], [Li(3\(s\))\(\cdot\)][H\(\cdot\)], etc.).

The diabatic configuration basis functions were then utilized to expand wavefunctions of the diabatic states. However, due to the non-orthogonality of the molecular orbitals between different fragments, the overlap integrals between diabatic configuration functions \(\Psi_i\),  \(S_{ij} = \langle \Psi_i|\Psi_j \rangle\),  should be evaluated. It also introduced additional complexity in computing the Hamiltonian matrix elements \(H_{ij} = \langle \Psi_i|\hat H^\text{ele}|\Psi_j \rangle\). We adopted the previously reported non-orthogonal configuration interaction treatment \cite{Thom2009, Sundstrom2014} and its recent extensions \cite{Straatsma2020, Straatsma2022} to address these challenges, which are detailed in the Supplementary Material. The resulting overlap matrix, \(\mathbf S\), and the diabatic potential energy matrix (DPEM), \(\mathbf H\), were integrated to form an effective model Hamiltonian for the concerned states of the system, which can be solved for approximately deriving the adiabatic or diabatic states. 

Moreover, a diabatic state was formed with a set of diabatic configuration basis functions. Certain matrix elements in \(\mathbf H\) and \(\mathbf S\), associating with the basis functions used for constructing a specific diabatic state, were picked out to form a subspace of the Hamiltonian. This subspace was represented by \(\mathbf{H}^\text{block}_p\) and \(\mathbf{S}^\text{block}_p\), with subscript denoting the diabatic state. The wavefunctions of these selected bases in the subspace were denoted as \( \Psi_i^{(p)} \) (\( i = 1, \ldots, N \)). The target diabatic state was then expressed as a linear combination of these basis functions, 
\begin{equation}
    \Phi_p^{\text{dia}} = \sum_{i=1}^N c_i^{(p)} \Psi_i^{(p)},
    \label{eq:diab_f}
\end{equation}
where the coefficients \{\( c_i^{(p)} \)\} were obtained by solving the generalized eigenvalue problem within the truncated non-orthogonal subspace:  
\begin{equation}
    \mathbf{H}^\text{block}_p \mathbf{C}_p = \mathbf{S}^\text{block}_p \mathbf{C}_p \mathbf{E}_p,
    \label{eq:eig_prob}
\end{equation} 
Here, \( \mathbf{C}_p \) is the eigenvector matrix, and \( \mathbf{E}_p \) is the diagonal matrix of eigenvalues. The diabatic wavefunction was constructed with Eq. \ref{eq:diab_f}, with combination coefficients being a selected eigenvector \( \mathbf{c}_k{^{(p)}} \) (index \(k\) is usually corresponding to the lowest eigenvalue).  

With constructed diabatic states, \(\{\Phi_p^{\text{dia}}\} \), a compact diabatic representation of the system was formed, with a contracted DPEM (\(\mathbf{H}^{\text{dia}}\)), as well as the corresponding overlap matrix (\(\mathbf{S}^{\text{dia}}\)), which were given by
\begin{equation}
    \mathbf H^{\text {dia}} = \mathbf U^\mathrm T \mathbf H \mathbf U, 
    \text{ and } 
    \mathbf S^{\text {dia}} = \mathbf U^\mathrm T \mathbf S \mathbf U, 
\end{equation}
where \(\mathbf U \) was constructed from the set of selected eigenvectors \( \mathbf{c}_k {^{(p)}} \) corresponding to specific diabatic states. The dimension of each eigenvector was expanded to match the total number of diabatic configuration basis functions. This dimensional expansion was achieved by assigning nonzero coefficients to the basis functions contributing to \( \mathbf{c}_k{^{(p)}} \), while the coefficients for all other basis functions were set to zero. 
Practically, the non-orthogonality of the contracted effective Hamiltonian \( \mathbf{H}^{\text{dia}}\) brought about inconvenience in further applications, various schemes were suggest previously to implementing the diabatic state orthogonalization, such as the L{\"o}wdin or Schmidt orthogonalization, resulting in the Hamiltonian, \(\mathbf{H}^\text{dia-o}\). 

\subsection{\label{sec:level2}Diabatic-to-adiabatic transformation }
Adiabatic states could be derived from diabatic states through a simple diagonalization:
\begin{equation}
    \mathbf{H}^{\text{dia}} \mathbf{L} = \mathbf{S}^{\text{dia}} \mathbf{L} \mathbf{E}^\text{adia},
    \label{eq:dta}
\end{equation}
where \(\mathbf L\) is the diabatic-to-adiabatic (DTA) transformation matrix, and the elements in the diagonal eigenvalue matrix \(\mathbf{E}^\text{adia}\) are the potential energies of the adiabatic states.
Alternatively, adiabatic states could be reproduced through a diagonalization process in the orthogonalized diabatic space, similar to Eq. \ref{eq:dta}, but using the Hamiltonian in its orthogonalized form, thus eliminating the need for an overlap matrix.
The non-adiabatic couplings (\(\boldsymbol d_{mn}\)) between adiabatic state \(m\) and \(n\) could be evaluated through the DTA transformation in the orthogonalized diabatic space as follows:
\begin{equation}
	\boldsymbol d_{mn} = \frac{\boldsymbol{l}^\mathrm{T}_m \nabla_\mathbf R \mathbf{H}^\text{dia-o} \boldsymbol{l}_n 
	}{E_n^\text{adia}-E_m^\text{adia}} ,
\end{equation} 
where \(E_n^\text{adia}\) and \(E_m^\text{adia}\) are the adiabatic energies associated with the adiabatic states \(m\) and \(n\), respectively. Moreover, \(\boldsymbol l_m\) and \(\boldsymbol l_n\) denote the corresponding columns of the DTA transformation matrix.

\section{\label{sec:level1} Results and discussion }
Usually, to accurately describe of the target adiabatic states, a slightly larger set of diabatic states should be constructed. In the demonstrative study of the four lowest \(^1\Sigma^+\) adiabatic states of LiH, six diabatic states were constructed by expanding them on a total of 23 CSFs, details are available in the Supplementary Material. The six diabatic states consists of one ionic state and five covalent states, denoted as
\(\Phi_{\mathrm{Li}^+\mathrm{H}^-}^\text{dia}\), 
\(\Phi_{\mathrm{Li}(2s)\mathrm{H}}^\text{dia}\), 
\(\Phi_{\mathrm{Li}(2p)\mathrm{H}}^\text{dia}\), 
\(\Phi_{\mathrm{Li}(3s)\mathrm{H}}^\text{dia}\), 
\(\Phi_{\mathrm{Li}(3p)\mathrm{H}}^\text{dia}\), and
\(\Phi_{\mathrm{Li}(3d)\mathrm{H}}^\text{dia}\).

The diabatic state \(\Phi_{\mathrm{ Li}^+\mathrm{H}^-}^\text{dia}\) represents an ionic configuration, while another ionic state \(\Phi_{\mathrm{ Li}(2s)\mathrm{H}}^\text{dia}\) is disregarded due to its unphysical electron distribution, which is insignificant in all adiabatic states under investigation. The remaining diabatic states fall under the category of covalent states. Apart from the two core electrons, the third electron of the Li atom can occupy either valence or Rydberg orbitals, such as Li(2\(s\)), Li(2\(p\)), Li(3\(s\)), Li(3\(p\)) and Li(3\(d\)), thereby forming various distinct diabatic states. To ensure the energy variational principle for diabatic states with the same symmetry, the Li(2\(s\))H and Li(3\(s\))H states were diagonalized, as well as the Li(2\(p\))H and Li(3\(p\))H states.

\subsection{\label{sec:level2}Validation }

\begin{figure}[htbp]
    \centering
    \includegraphics[width=3in]{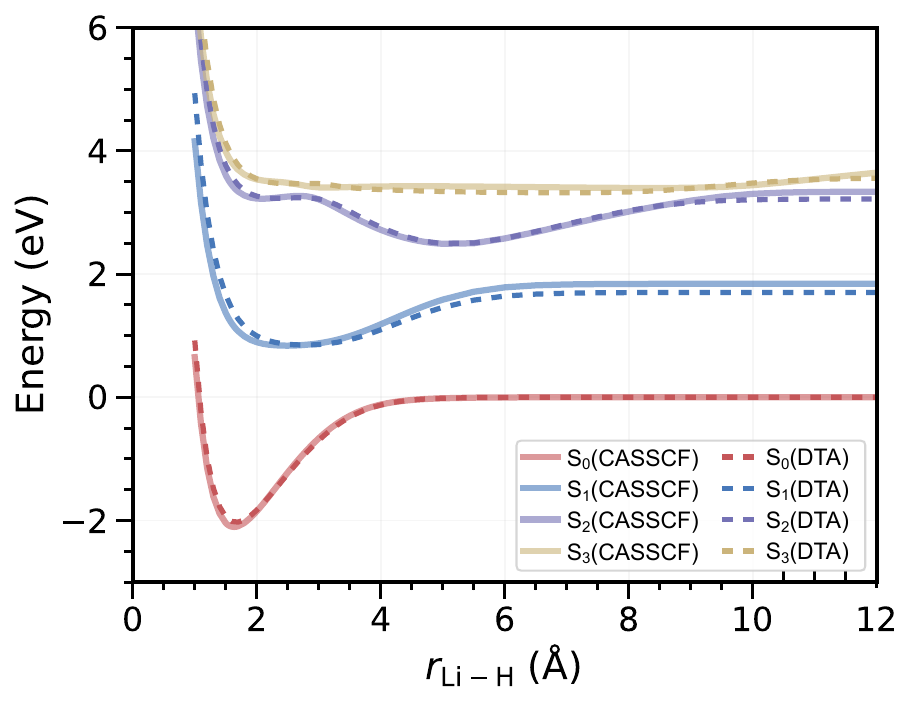}
    \caption{The four lowest \(^1\Sigma^+\) adiabatic potential energy curves generated by diabatic-to-adiabatic (DTA) transformation with our approach (dashed lines) and corresponding SA(4)-CASSCF(2,9) results (solid lines).}
    \label{fig:adia_tot}
\end{figure}

The suitability of the selected CSF set and the effectiveness of the proposed diabatic state construction method were validated against the CASSCF method, through characterizing the four lowest \(^1\Sigma_+\) adiabatic states (S\(_0\)-S\(_3\)). Examination of the adiabatic PESs illustrated in Fig.~\ref{fig:adia_tot}, as well as the energetic and geometric parameters summarized in Fig.~\ref{supp-tab:energy}, proves that the critical features of the adiabatic states identified by the CASSCF calculations are quantitatively preserved in the results obtained from our method. 
For the ground-state (S\(_0\)), the dissociation energy of 2.03 eV from our approach is consistent with the value of 2.10 eV from CASSCF. The equilibrium bond length of 1.67  {\AA} also closely matches the CASSCF value of 1.64 {\AA}. Additionally, the flat bottom of the PES for state S\(_1\) is resolved using our method, showing a well depth of 0.87 eV, which aligns with 1.00 eV from CASSCF result. The flat bottom complicates the precise identification of the equilibrium bond length, leading to a value of 2.77 {\AA} in the present method, slightly larger than the CASSCF result of 2.48 {\AA}. 

Both S\(_2\) and S\(_3\) electronic states exhibit obvious Rydberg characteristics in their wavefunctions, increasing the complexity of the PES landscape, which were observed at the CASSCF theoretical level and were successfully reproduced with our method. The most distinctive signature of the S\(_2\) state manifests as an asymmetric double-well potential, characterized by two equilibrium geometries at 2.32 {\AA} and 5.21 {\AA} and an energy barrier at 2.71 {\AA}, which are in agreement with CASSCF reference values (minima: 2.08 {\AA} and 5.19 {\AA}; transition state: 2.69 {\AA}). The energy difference between the two well bottoms is determined as 0.74 eV in our method, in consistence with the CASSCF value of 0.74 eV.
Both our method and CASSCF characterize an extensive basin-shaped potential well in the PES of the S\(_3\) state, spanning nuclear separations from approximately 2.0 {\AA} to 9.5 {\AA}. 
Within this region, the S\(_3\) state undergoes two avoided crossings with the S\(_2\) state. The first avoided crossing occurs at \( r_{\text{\ce{Li-H}}} = 2.65 \,  \text{\AA}\), near the transition point of the S\(_2\) state, while the second is located at \( r_{\text{\ce{Li-H}}} = 8.96 \,  \text{\AA}\), close to the end of the basin (2.83 {\AA} and 9.92 {\AA} respectively in CASSCF results).

\subsection{\label{sec:level2}The constructed diabatic states}

\begin{figure}[htbp]
	\centering
	\includegraphics[width=3in]{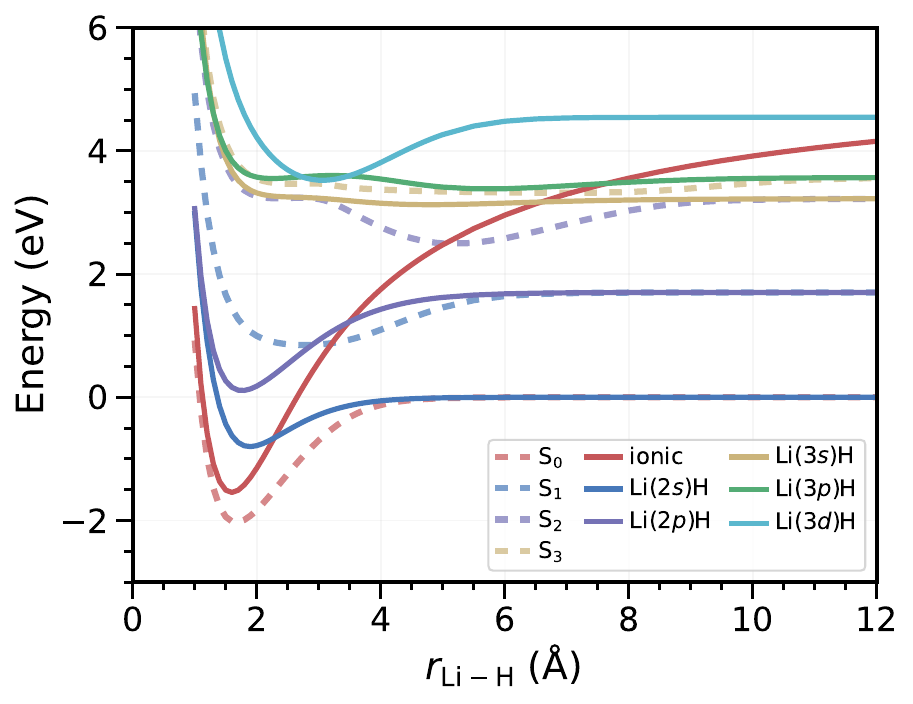}
	\caption{The potential energy curves for six diabatic states (solid lines), and the resulting adiabatic potential energy curves (dashed lines).}
	\label{fig:dpes}
\end{figure}

Using invariant molecular orbitals brings convenience in characterizing the diabatic wavefunctions, and hence in disclosing the evolution of the wavefunctions along with the geometrical propagation, particularly before these wavefunctions are mixed by orthogonalization. Fig.~\ref{fig:dpes} depicts the potential energy curves of the six non-orthogonalized diabatic states. The natural orbitals for each diabatic state, as well as the occupation numbers are shown in Supplementary Material.

The ionic state Li\(^+\)H\(^-\) is characterized by a deep potential well centered at a \ce{Li-H} distance of 1.60 {\AA}. Beyond this minimum, the potential energy rises sharply as the bond stretches, driven by the electrostatic attraction between the Li\(^+\) and H\(^-\) ions. Notably, the two valence electrons of LiH predominantly localize in the H(1\(s\)) orbital across all bond lengths, as evidenced by the shapes of natural orbitals and their occupation numbers shown in Fig.~\ref{supp-fig:ionic_no}. In contrast, the five covalent states exhibit significantly shallower or nearly absent potential wells. Among these states, the energy ordering generally follows the sequence, Li(2\(s\))H, Li(2\(p\))H, Li(3\(s\))H, Li(3\(p\))H and Li(3\(d\))H. Moreover, these states cross the ionic state in the same order. 
The Li(2\(s\))H state exhibits a potential well with a depth of 0.80 eV, centered at \( r_{\text{\ce{Li-H}}} = 1.89 \, \text{\AA} \). This diabatic state originates from the valence bonding of Li(2\(s\)) and H(1\(s\)) orbital, as shown in Fig.~\ref{supp-fig:2s_no}, but its anti-bonding characteristic weakens the bond strength. Furthermore, it crosses the ionic state at \( r_{\text{\ce{Li-H}}} = 2.28 \, \text{\AA} \), triggering adiabatic electronic configuration rearrangements, which will be analyzed in the following content. 
The Li(2\(p\))H state is energetically close to the Li(2\(s\))H state at short \ce{Li-H} distances, suggesting the possibility of \(sp\) hybridization in the bonding process. It displays a slightly deeper potential well than Li(2\(s\))H, centered at \( r_{\text{\ce{Li-H}}} = 1.76 \, \text{\AA} \), with the depth of 1.59 eV. This can be attributed to the directional nature of the Li(2\(p\)) orbital compared to the spherical Li(2(\(s\))) orbital, which enhances bonding overlap and strengthens the bond. Consequently, this not only yields a deeper well but also shortens the minimum bond length. However, the higher energy of the Li(2\(p\))H state shifts its crossing point with the ionic state to a longer interatomic distance of 3.48 {\AA}.  

The three higher-lying covalent states are constructed as Rydberg states. Natural orbital analyses reveal their distinctions from the lower-lying valence states, primarily arising from the spatial delocalization and nodal structures of the Rydberg orbitals. Specifically, the Li(3\(s\))H state exhibits destabilization due to weakened bonding interactions, originated from the simultaneous overlap of the H(1\(s\)) orbital with both positive and negative phase regions on either side of the nodal structure of the Li(3\(s\)) orbital, as shown in Fig.~\ref{supp-fig:3s_no}, leading to partial cancellation of bonding contributions. Meanwhile, during bond stretches, the H(1\(s\)) orbital engages in alternating bonding interactions with successive nodal regions of the Li(3\(s\)) orbital, transitioning from inner to outer nodal domains.
Similar characteristic is observed in the Li(3\(p\))H state according to Fig.~\ref{supp-fig:3p_no}. Consequently, such states exhibit remarkably flat potential energy curves throughout the internuclear distance range. In contrast, lacking radial nodes in the Li(3\(d\)) orbital alleviates the multi-region phase cancellation, which is displayed in Fig.~\ref{supp-fig:3d_no}, so that the Li(3\(d\))H PES qualitatively resembles low-lying covalent states, albeit with a substantially elongated equilibrium bond length \(r_{\text{\ce{Li-H}}}=3.06\,  \text{\AA}\) induced by the more diffusive Rydberg orbital.

\subsection{\label{sec:level2}Characterization of adiabatic states }

\begin{figure}[htbp]
	\centering
	\includegraphics[width=3.3in]{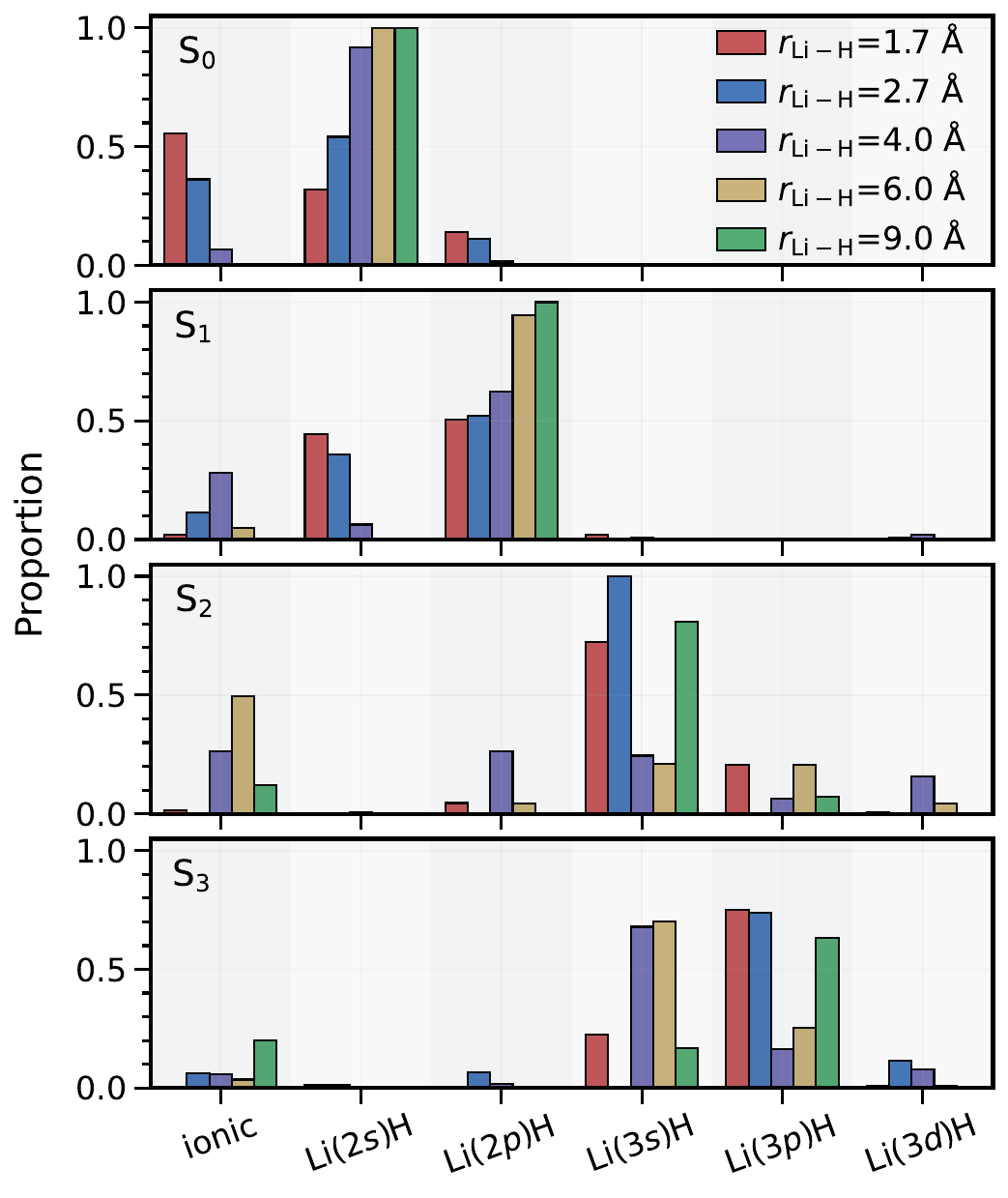}
	\caption{The proportions of diabatic states in the four adiabatic states at some selected \ce{Li-H} bond lengths.}
	\label{fig:co_bar}
\end{figure}

An in-depth analysis of adiabatic state wavefunction along with the bond dissociation has been performed using the calculated diabatic state contributions to these adiabatic states, as detailed in Supplementary Material. For brevity, key geometric configurations and their corresponding diabatic state proportions are summarized with histogram plots shown in Fig.~\ref{fig:co_bar}.

In general, the competition between ionic and covalent states gives rise to the formation of the potential well in each adiabatic PES. The Li(2\(s\))H state and the ionic state jointly shape the behavior of the S\(_0\) adiabatic state, which is shown in Fig.~\ref{fig:co_bar}, Fig.~\ref{supp-fig:co_s0}. The crossing point of the proportions of the ionic and the covalent states in S\(_0\) locates at \( r_{\text{\ce{Li-H}}} = 2.31 \,  \text{\AA}\). While the ionic state prevails in structures when \( r_{\text{\ce{Li-H}}}\) is shorter than \(2.31 \,  \text{\AA}\), the covalent state becomes predominant when nuclear separation is beyond \(4.0 \,  \text{\AA}\), leading to a purely covalent S\(_0\) state towards the dissociation limit.

As shown in Fig.~\ref{fig:co_bar}, Fig.~\ref{supp-fig:co_s1}, wavefunction analysis of the S\(_1\) state indicates that the Li(2\(p\))H diabatic state dominates at nuclear separations beyond \(6\,  \text{\AA}\). At shorter distances, the S\(_1\) state exhibits non-negligible contributions from both the Li(2\(s\))H and ionic states. As the two atoms approach, the contribution from Li(2\(s\))H increases and eventually surpasses that of the Li(2\(p\))H state in S\(_1\) state, reflecting strong hybridization between the lithium 2\(s\) and 2\(p\) orbitals. Around a bond length of 4 {\AA}, ionic characteristic becomes more pronounced, emerging from the interplay between the Li(2\(p\))H and ionic state. The orbital hybridization between Li(2\(s\))H and Li(2\(p\))H, as well as the resulting coupling with the ionic state, plays a key role in the formation of the broad potential well. The hybrid Li(2\(ps\))H state displays a shallower potential well compared to the Li(2\(p\))H state, with its minimum located at \( r_{\text{\ce{Li-H}}} = 2.38 \,  \text{\AA}\), as shown in Fig.~\ref{supp-fig:sp}. Meanwhile, it mixes with the ionic state in the vicinity of \( r_{\text{\ce{Li-H}}} = 4 \,  \text{\AA}\), cooperatively contributing to the formation of the broad well in this region. Further analyses of hybrid diabatic states based on their diabatic PESs and natural orbitals are provided in the Supplementary Material Fig.~\ref{supp-fig:2sp_no}, Fig.~\ref{supp-fig:2ps_no}.

Compared to the S\(_0\) and S\(_1\) states, the S\(_2\) and S\(_3\) states exhibit more complex electronic structures owing to the involvement of Rydberg orbitals. According to Fig.~\ref{fig:co_bar}, Fig.~\ref{supp-fig:co_s2}, near an internuclear distance of \(2.71\,  \text{\AA}\), the S\(_2\) state is primarily characterized by the Li(3\(s\))H configuration. As the two atoms are further separated, the dominant contribution to S\(_2\) shifts to the ionic state, and finally returns to the Li(3\(s\))H configuration. The covalent interaction between the Li(3\(s\)) or Li(3\(p\)) and H(1\(s\)) electrons drives the formation of the first potential energy well in the S\(_2\) state at \( r_{\text{\ce{Li-H}}} = 2.32 \,  \text{\AA}\). Additionally, the ionic state contributes significantly over a wide range of bond lengths from about 3 {\AA} to 10 {\AA}, which results in the formation of the second potential well.  

Similarly, the S\(_3\) state undergoes analogous transformations with increasing internuclear distance in Fig.~\ref{fig:co_bar}, Fig.~\ref{supp-fig:co_s3}. The dominant contribution evolves from the Li(3\(p\))H state to the Li(3\(s\))H configuration, and eventually reverts back to Li(3\(p\))H. While the ionic state participates in the competition within the S\(_3\) state, its influence remains secondary. It reaches a maximum contribution at \(8.5\,  \text{\AA}\) where it stabilizes a shallow potential well. Sharp variation of the diabatic state proportion curves of S\(_2\) and S\(_3\) is observed near 2.71 {\AA}, which is illustrated in Fig.~\ref{supp-fig:co_s2}, Fig.~\ref{supp-fig:co_s3}. These abrupt changes reflect instantaneous switching between electronic configurations around thus formed avoided crossings, indicative of strong nuclear-electron coupling during geometric evolution. The Li(3\(d\))H plays a crucial role in shaping the curves of states S\(_2\) and S\(_3\) around the avoided crossing point at \(2.71\,  \text{\AA}\), suggested by the abrupt increase of its proportion in this region. Therefore, it is reasonable to consider Li(3\(d\)) acting as a polarization function that enhances bonding interaction between the \(p\) orbital of Li and  H(1\(s\)) orbitals, particularly in the S\(_2\) and S\(_3\) states.

\subsection{\label{sec:level2}Orthogonalization of diabatic states}
\begin{figure}[htbp]
    \centering
    \captionsetup[subfigure]{labelformat=empty}
    
    \begin{subfigure}[b]{3in}
        \includegraphics[width=3in]{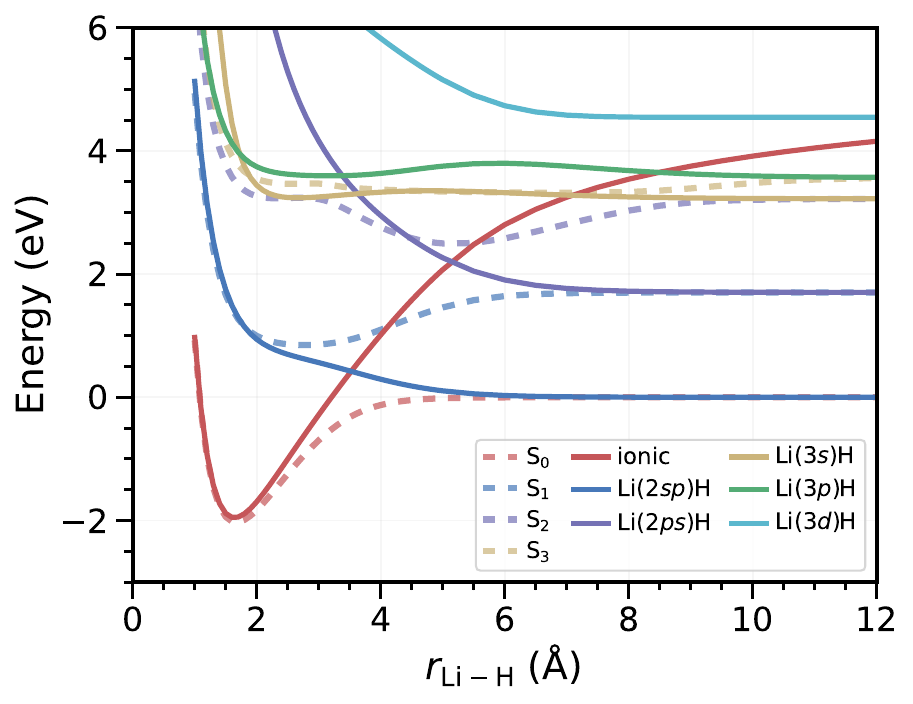}
        \caption*{(a)}
        \label{fig:ortho_sub1}
    \end{subfigure}
    \begin{subfigure}[b]{3in}
        \includegraphics[width=3in]{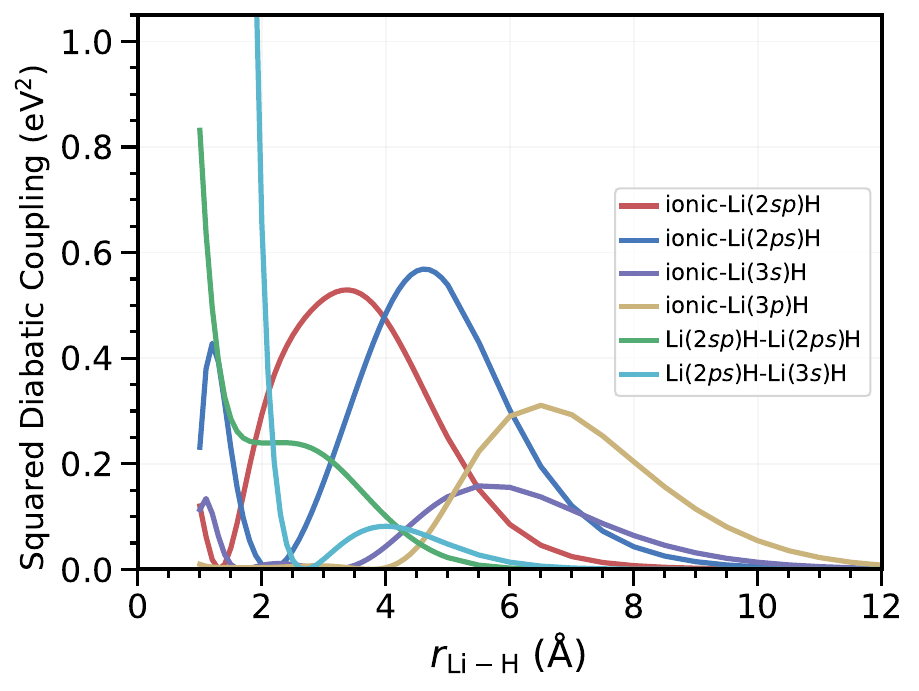}
        \caption*{(b)}
        \label{fig:ortho_sub2}
    \end{subfigure}
    
    \caption{(a) The orthogonal diabatic states using Schmidt orthogonalization, and (b) squared diabatic couplings between selected pairs of these diabatic states}
    \label{fig:ortho}
\end{figure}

Orthogonalized diabatic wavefunctions offer advantages for analyzing diabatic Hamiltonian properties and facilitating non-adiabatic dynamics simulations. In this work, the ionic, Li(2\(s\))H and Li(2\(p\))H states underwent preprocessing prior to the orthogonalization procedure. Firstly, The ionic state was linearly mixed with the Li(2s)H and Li(2p)H state according to their respective overlap integrals \(S_{\mathrm {ionic}/2s}\) and \(S_{\mathrm {ionic}/2p}\),
\begin{equation}  
    \Phi_{\mathrm{Li}^+\mathrm{H}^-}^\text{dia} = c'(\Phi_{\mathrm{Li}^+\mathrm{H}^-}^\text{dia} + S_{\mathrm {ionic}/2s}\Phi_{\mathrm{Li}(2s)\mathrm{H}}^\text{dia}+ S_{\mathrm {ionic}/2p}\Phi_{\mathrm{Li}(2p)\mathrm{H}}^\text{dia}),
    \label{eq:orthogon}  
\end{equation}  
where \(c'\) is the normalization factor. This preprocessing modifies the curvature of the ionic potential well, bringing the curve closer to the S\(_0\) state PES near the equilibrium geometry.
A similar mixing strategy addressed \(sp\)-hybridization in short-bond regions, generating two distinct hybrid states denoted as  Li(2\(sp\))H and Li(2\(ps\))H.
 
The orthogonalization procedure utilized the reconstructed ionic state as the reference, followed by conducting sequential Schmidt orthogonalization on the remaining states in the following order: Li(\(2sp\))H, Li(\(2ps\))H, Li(\(3s\))H, Li(\(3p\))H, and Li(\(3d\))H. Fig.~\ref{fig:ortho}(a) presents the resultant orthogonal diabatic PESs. These orthogonalized PESs show a high degree of consistency with previous findings\cite{Grofe2017}, which offers strong validation for our DDSC method. Based on these diabatic states, we computed the corresponding diabatic couplings directly, with the results displayed in Fig.~\ref{fig:ortho}(b). As expected, the most significant diabatic coupling occurs near the crossing point of the two diabatic states shown in Fig.~\ref{fig:ortho}(a). This further confirms the validity of our diabatic state definition, as in a well-defined diabatic representation, the largest couplings between two diabatic states are predicted to occur at their crossing points.

\subsection{\label{sec:level2}The non-adiabatic couplings of adiabatic states}

\begin{figure}[htbp]
	\centering
	\includegraphics[width=3in]{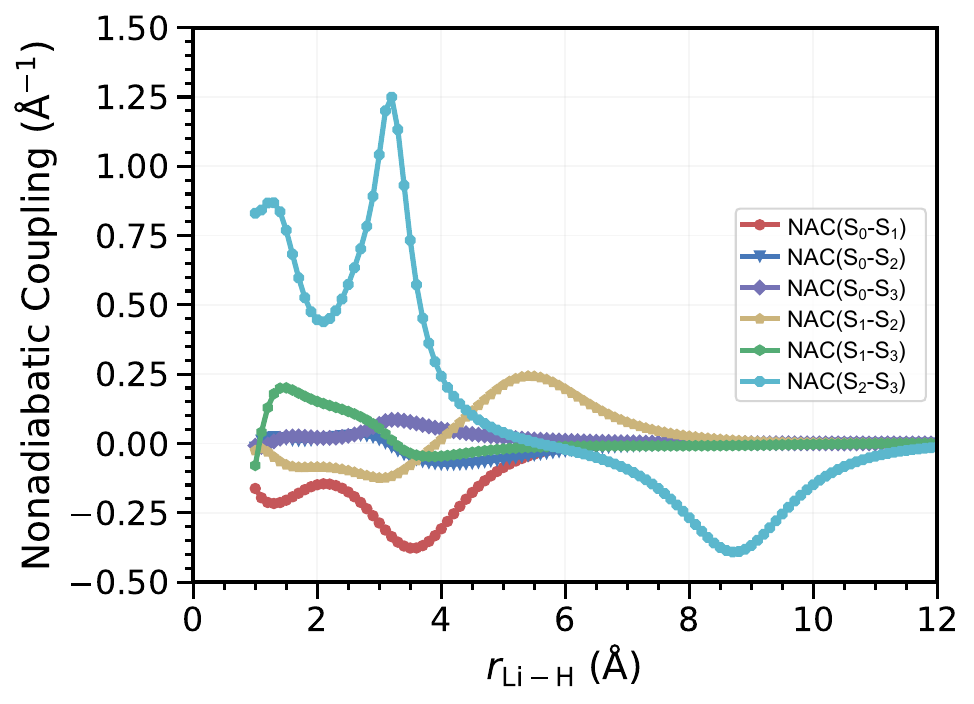}
	\caption{The non-adiabatic couplings between these adiabatic states derived from the DTA transformation}
	\label{fig:nac}
\end{figure}

The non-adiabatic couplings between the adiabatic states along the reaction coordinate were calculated via the DTA transformation, which are presented in Fig.~\ref{fig:nac}. The results align with our previous discussion about the energy proximity dependence, strongest couplings emerge between adjacent energy states, particularly in the S\(_0\)-S\(_1\), S\(_1\)-S\(_2\), S\(_2\)-S\(_3\) pairs.
Moreover, the singular coupling topology is reproduced, which is shown as the distinctive sharp cusp-like NAC profile of S\(_2\)-S\(_3\), demonstrating a remarkable non-adiabatic feature between these two states. 
While such singular points pose challenges for conventional adiabatic-state based approaches due to numerical divergences, our scheme enables their identification through a transformation based on well-defined diabatic states. Thus, this method has broader applicability in the excited-state molecular dynamic simulation.

In summary, this work introduces a DDSC method, where all diabatic states can be constructed at the same foot by employing FL-SC MOs. 
Imposing invariant fragment orbitals across diabatic states and geometries enhances computational efficiency while preserving accuracy with a compact diabatic basis set. 
The application on the LiH molecule validates the method's effectiveness in constructing diabatic states and computing diabatic couplings which can capture key features of its four lowest adiabatic states. 
This method is particularly suitable for studying charge transfer in weakly interacting large-scale systems. However, strongly interacting systems pose challenges due to substantial orbital distortions that occur when fragments are in proximity. In such cases, the inclusion of more diabatic bases becomes inevitable. Despite these limitations, the current framework remains advantageous for applications dominated by weak fragment interactions. Future extensions that integrate adaptive configuration selection and orbital optimization strategies within the present diabatic framework could broaden its applicability to more complex interaction situations. Additionally, systematic improvements can be achieved by expanding the diabatic basis or incorporating dynamic correlations via perturbation theory or density functional theory.

\section*{SUPPLEMENTARY MATERIAL}
See supplementary material for further computational details, and additional results related to Section III presented as tables and figures.

\begin{acknowledgements}
This work was supported by the National Key Research and Development Programs of China (2023YFA1506902) and the National Natural Science Foundation of China NSF (Grand Nos. 22073041, 22373040). We thank the High Performance Computing Center of Nanjing University for computational resources.
\end{acknowledgements}

\clearpage
\bibliography{reference}

\end{document}